\documentclass{article}

\usepackage{arxiv}
\usepackage{multirow}
\usepackage{listings} 
\usepackage{amssymb}
\usepackage{amsmath}

\usepackage[utf8]{inputenc} 
\usepackage[T1]{fontenc}    
\usepackage{hyperref}       
\usepackage{url}            
\usepackage{booktabs}       
\usepackage{amsfonts}       
\usepackage{nicefrac}       
\usepackage{microtype}      
\usepackage{lipsum}		
\usepackage{graphicx}
\usepackage{natbib}
\usepackage{doi}
\usepackage{algorithmic}
\usepackage{textcomp}
\usepackage{xcolor}
\usepackage{multirow}
\usepackage{lscape}
\usepackage{pifont}

\usepackage{array}
\usepackage{listings} 
\usepackage{easylist} 
\usepackage[most]{tcolorbox}
\usepackage{inconsolata}
\usepackage{tikz}
\usepackage{gensymb}
\usepackage{authblk}
\usetikzlibrary{positioning,decorations.pathreplacing,fit}

\newtcblisting[auto counter]{sexylisting}[2][]{sharp corners, fonttitle=\bfseries, colframe=gray,  listing only, listing options={linewidth=\columnwidth,breaklines=true,basicstyle=\scriptsize,language=java, commentstyle = \color{purple} \scriptsize, escapeinside={<@}{@>}},  title=Listing \thetcbcounter: #2, #1}
\def\BibTeX{{\rm B\kern-.05em{\sc i\kern-.025em b}\kern-.08em
    T\kern-.1667em\lower.7ex\hbox{E}\kern-.125emX}}
\newcommand*\samethanks[1][\value{footnote}]{\footnotemark[#1]}

\title{Chat2Code: Towards conversational concrete syntax for model specification and code generation, the case of smart contracts}

\author[1\thanks{These authors contributed equally to this work}]{Ilham Qasse} 
\author[2\samethanks]{Shailesh Mishra}
\author[1,3]{Mohammad Hamdaqa}
\affil[1]{Department of Computer Science, Reykjavik University, Reykjavik, Iceland \authorcr
  \{\tt ilham20, mhamdaqa\}@ru.is}
\affil[2]{Department of Electrical Engineering, Indian Institute of Technology Kharagpur, Kharagpur, India \authorcr
  \tt mshailesh0511@iitkgp.ac.in}
\affil[3]{Department of Computer and Software Engineering, Polytechnique Montreal, Montreal, Canada \authorcr
  \tt mhamdaqa@polymtl.ca}
\date{}
\renewcommand{\shorttitle}{\textit{arXiv} Template}

\hypersetup{
pdftitle={Chat2Code: Towards conversational concrete syntax for model specification and code generation, the case of smart contracts},
pdfsubject={q-bio.NC, q-bio.QM},
pdfauthor={Ilham Qasse, Shailesh Mishra},
pdfkeywords={Model-driven Engineering, Automatic Code Generation, Chatbots, Smart Contracts, Blockchain, Natural Language Processing},
}

\begin{document}
\maketitle

\begin{abstract}
The revolutionary potential of automatic code generation tools based on Model-Driven Engineering (MDE) frameworks has yet to be realized. Beyond their ability to help software professionals write more accurate, reusable code, they could make programming accessible for a whole new class of non-technical users. However, non-technical users have been slow to embrace these tools. This may be because their concrete syntax is often patterned after the operations of textual or graphical interfaces. The interfaces are common, but users would need more extensive, precise and detailed knowledge of them than they can be assumed to have, to use them as concrete syntax.

Conversational interfaces (chatbots) offer a much more accessible way for non-technical users to generate code. In this paper, we discuss the basic challenge of integrating conversational agents within Model-Driven Engineering (MDE) frameworks, then turn to look at a specific application: the auto-generation of smart contract code in multiple languages by non-technical users, based on conversational syntax. We demonstrate how this can be done, and evaluate our approach by conducting user experience survey to assess the usability and functionality of the chatbot framework.
\end{abstract}



\section{Introduction}
Model-Driven Engineering (MDE) accelerates the development of complex software by raising its abstraction level \cite{brambilla2017model}. One main application for MDE is automatic code generation from system designs \cite{niaz2005automatic}. This usage has drawn attention because it produces more accurate, less error-prone, more reusable code that is easier to maintain than manually written code \cite{niaz2005automatic}.  Over the past two decades, many researchers \cite{tran2018lorikeet,tolvanen2016model,hamdaqa2020icontractml,gharaat2021alba, nguyen2015frasad, ciccozzi2017model} have successfully used MDE to build frameworks for auto-generating code and applications in domain-specific fields like the Internet of Things, smart contracts, mobile applications, etc. 
Most MDE-based tools and development frameworks use textual or graphical interfaces as the concrete syntax for the developed languages. This makes it easy for domain experts to specify models, as it uses concepts they already know. However, this isn't as helpful for non-technical users, for at least three reasons. First, they need to learn both the interfaces and the underlying syntax of the newly introduced languages \cite{perez2019towards,perez2019flexible}. Second, these interfaces provide delayed feedback to the user, if any. This frustrates new non-technical users and makes them quickly abandon the frameworks \cite{perez2019towards,perez2019flexible}. 
Finally, while many of the MDE development frameworks try to bridge the gap between the different software development stakeholders, interaction and collaboration between them are normally limited to providing them with different interfaces, not introducing mechanisms to facilitate active interaction and collaboration\cite{perez2019towards,perez2019flexible}.
In this paper, we demonstrate how these problems may be circumvented using chatbots (conversational interfaces) as a concrete syntax for domain-specific languages. This syntax can facilitate the specification of domain models, and auto-generate code from these models. 


A chatbot is a software application that uses Natural Language (NL) conversation to conduct conversations (online chat) with users \cite{cahn2017chatbot,abdellatif2020challenges}.  Chatbots are common information-gathering interfaces. There are several advantages in using them as a concrete syntax or a layer on top of a domain-specific language: (i) they are easy to integrate as they do not require any installation, (ii) the NL interface eases the learning curve and reduces the need for training in the notations and representations of the concrete syntax of domain-specific languages, compared to other modelling tools (e.g. deployed within eclipse) \cite{perez2019towards,perez2019flexible}, (iii) they provide direct and immediate feedback to the user, guiding the gradual and incremental construction of their application model, and (iv) they better suited for collaboration between different stakeholders, since they emulate natural interactions.
Nevertheless, when using chatbot or NL in general, users normally define incomplete or incorrect input in their conversation, especially non-technical users. In this case, we are required to detect and correct the user input to accurately generate model specifications. 
Moreover, to target a wider range of users and platforms, the concrete syntax must enable users to freely express their program ideas with the terminologies they are familiar with. For example when defining a numerical data type depending on the user (whether programmer or non-programmer ) it can be number, decimal, integer, float, etc. This mismatch between the terminologies must be considered in the chatbot, and it should be able to map between these terminologies. 
 Furthermore, the generated artefacts and model should be traced to user conversation, to detect and evaluate these artefacts and compare them to the user request. 

In this paper, we aim to address these challenges in order to auto-generate codes flexibly from conversational concrete syntax. To show the feasibility of our approach we implement a chatbot framework, iContractBot,  to design and develop smart contract codes from existing smart contracts reference meta-model.  
To summarize the most salient contributions of our research, we:
\begin{enumerate}
    \item investigate the usage of chatbots as a concrete syntax layer for modelling frameworks. 
    \item propose an approach to auto-generate platform-independent code from conversational syntax.
    \item provide an implementation of our methodology to generate smart contract code for multiple blockchain platforms.
    \item evaluate the generalizability of the chatbot based on real case studies.  
    \item conduct user experience survey to evaluate the functionally and usability of chatbot framework.
\end{enumerate}

The paper is organized as follows: 
Section \ref{RM} proposes a method for defining the conversational syntax for Domain-Specific Languages (DSL). Section
 \ref{IMP} demonstrates the study case implementation. Section \ref{EV} presents an evaluation of the paper's contributions. Limitations of the study and threats to validity are presented in Sections \ref{LIMR} and \ref{TH} respectively. Related work is covered in Section \ref{RW}.
Finally, a conclusion and future research directions are presented in Section \ref{CON}.

\section{Conversational Syntax for Domain-Specific Languages}
\label{RM}
The main goal of this paper is to demonstrate an approach for using chat conversations to generate code. To achieve this, we are following a model-based approach.  Given a model-driven development framework that can be used to generate code from models, the goal is to extend it to support conversational model specification. This is a multi-part challenge. The information-gathering flow of the chatbot must conform to the meta-model/abstract syntax of the modelling framework. Then NL input must be transformed through that chatbot to an instance model specification that can be used to generate code as output. We approach the conformance problem by constructing an intermediate layer between the meta-model and chatbot enabling the two to be reciprocally mapped. We solve the NL translation to a model instance using a 5-step transformation process.   

\subsection{Defining Chatbot Specifications}
\begin{figure}
\centerline{\includegraphics[scale = 0.8]{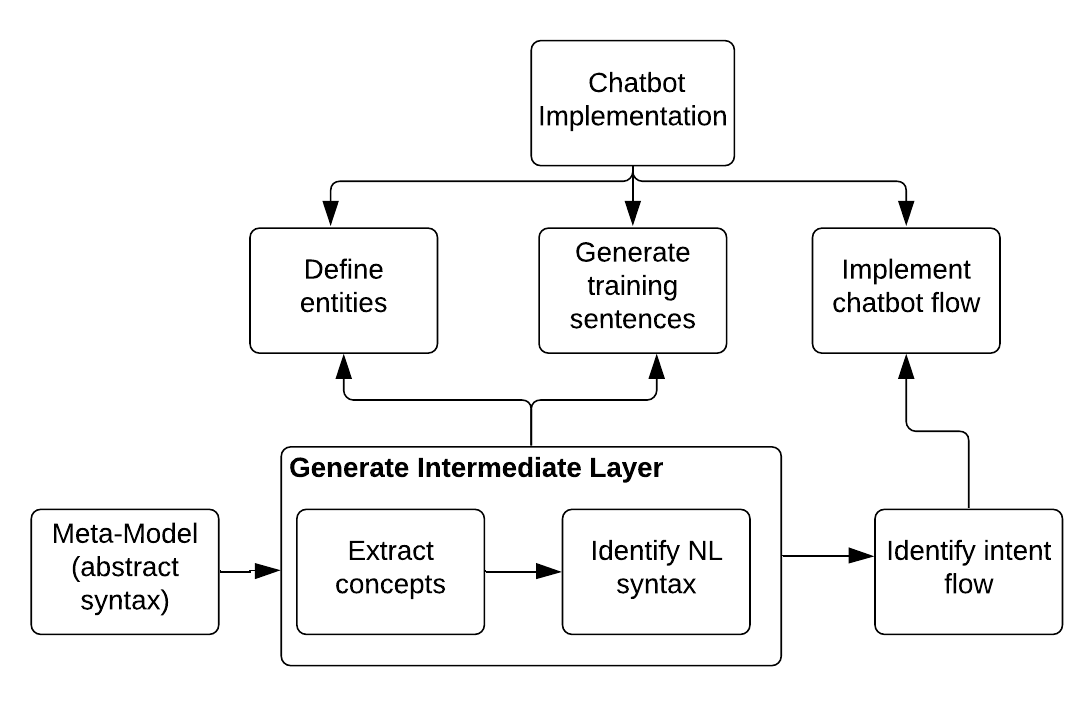}}
\caption{Research Methodology to generate chatbot specifications}
\label{fig:RM-1}
\end{figure}

Figure \ref{fig:RM-1} illustrates our approach for structuring chatbot interactivity based on existing meta-models. It consists of two main steps: constructing an intermediate layer that ensures meta-model conformity, and defining chatbot intent flow to gather instance model specifications.

\subsubsection{Generate Intermediate Layer} \label{IL}
We start by extracting the concepts of the language (vocabulary and taxonomy) which are captured by the meta-model/abstract syntax. Based on those concepts, we define a set of synonyms that describe natural language representations of concepts in an intermediate layer, between the abstract and concrete syntax. We also identify natural language synonyms in that layer for attributes from the meta-model and possible operations that can be applied to both concepts and attributes. Then we generate sets of training sentences for each concept and operation that can be used to capture user intent. This intermediate layer serves to link the abstract syntax with the conversational syntax, identifying how the elements are related, and mapping them to each other. 

\subsubsection{Chatbot Intent Flow}

\begin{figure} 
\resizebox{\linewidth}{!}{
\centerline{\includegraphics[scale = 0.6]{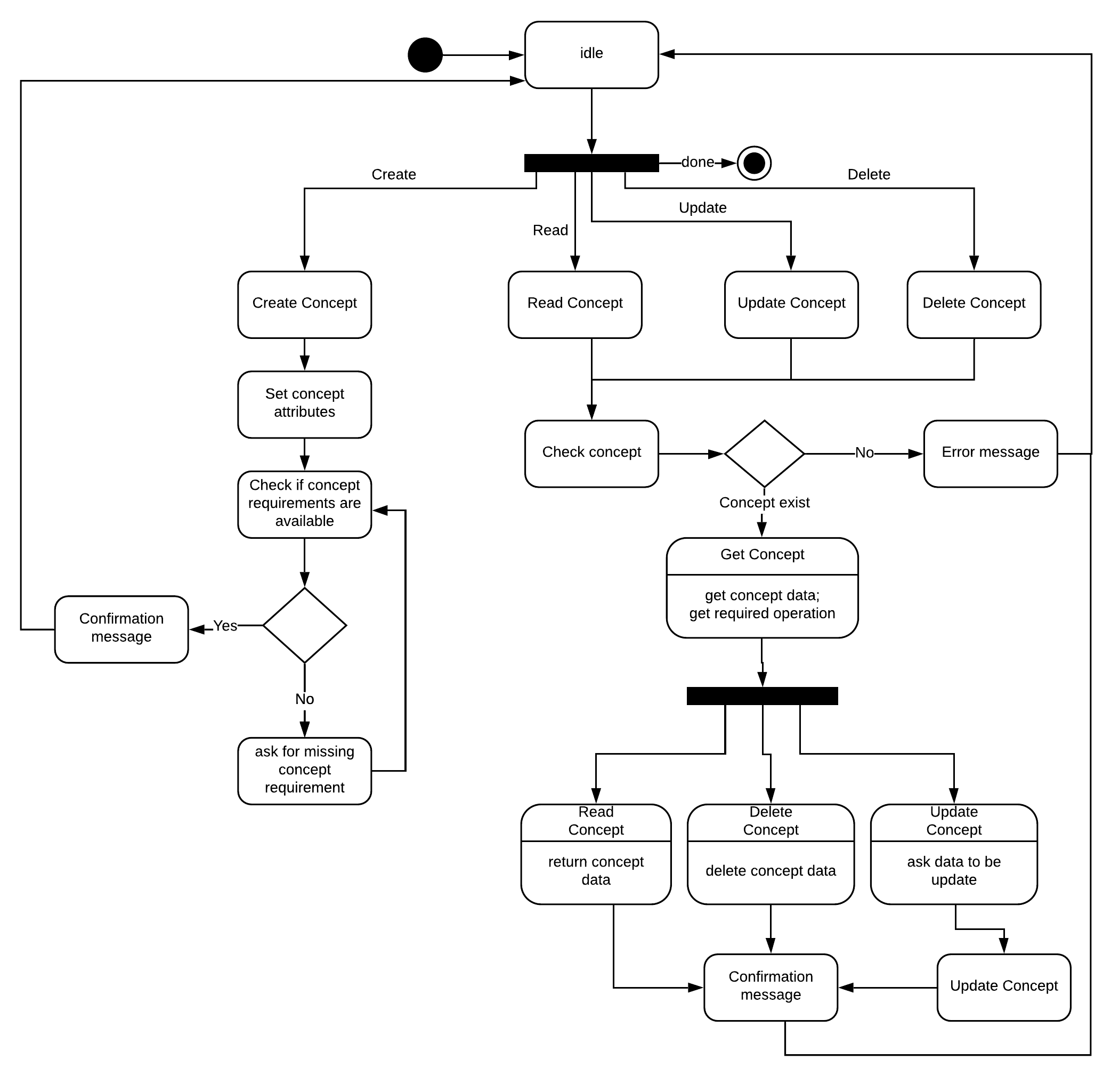}}}
\caption{Chatbot Intent Flow}
\label{fig:GINM}
\vspace{-5mm}
\end{figure}

Since we are targeting non-technical users, we need to direct them to efficiently articulate the concepts required to produce complete models. For this reason, the intent model driving conversation flow must conform tightly to meta-model concepts. Potential actions commonly used to interact with the meta-model must also be captured by conversation flow. We have used a statechart to represent the intent flow underlying this process. Each concept is represented as a state, with subsequent states representing CRUD operations (create, read, update and delete) that might be required in the user-chatbot conversation to validate input. Figure \ref{fig:GINM} shows this intent flow. Although only a definite set of operations has been used, the operations encompass all the actions that can be executed on a meta-model. Thus, these operations make the chatbot accurate and versatile at the same time.
The create operation is to create a model instance from the meta-model, while the operations read, update and delete queries an existing instance model. In the case of model instance query, as shown in Figure  \ref{fig:GINM}, the chatbot makes sure that the concept exists before reading, updating, or deleting a concept.  Rules can be enforced in the statechart to ensure that the flow is correct, and over several cycles, a complete model can be extracted from the user conversation. With the specification of this intent flow, and on the basis of entities and training sentences based on the intermediate layer, a chatbot can be implemented to gather information from the user that ultimately maps to the meta-model through the intermediate layer, allowing for the generation of a functional instance model, and the resulting code.

\subsection{Generating Model Specification}\label{AA}

\begin{figure}[t]
\centering
\includegraphics[scale = 0.7]{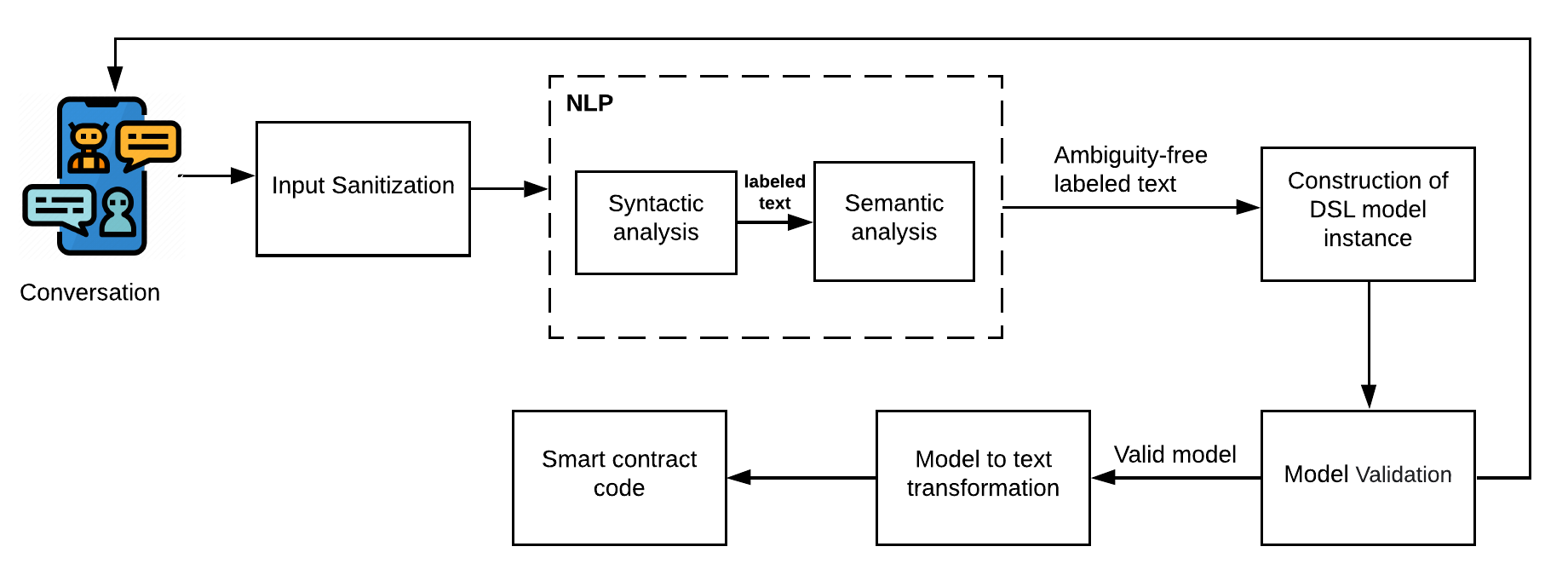}
\caption{Research Methodology to generate model specifications and code from conversational input}
\label{fig:RM-2}
\vspace{-5mm}
\end{figure}

Now that we have discussed our general approach for mapping the meta-model concepts to a chatbot intent model, we need to specify the process for taking NL input from the user and working it through this model-based conversational system, to generate instance model specifications and code from it. Figure \ref{fig:RM-2} demonstrates our approach to that problem. It involves the following 5 steps:
\subsubsection{Input Sanitization}  When using conversational agents or NL interfaces in general, users frequently provide incomplete or incorrect input, especially non-technical users. We have to detect deviations and correct user input to accurately generate model specifications. 
We use the Levenshtein Distance algorithm \cite{yujian2007normalized} to modify NL input based on similar existing data and confirm those modifications with the user. The Levenshtein Distance algorithm finds what is called the \emph{edit distance} between two strings. Edit distance infers the number of changes that would need to be made to String A so that it becomes equal to String B. Handling this kind of deviation and correction in the early stages of our process adds more flexibility to our approach.
\subsubsection{NLP Component} The NLP component is mainly used to detect and identify user intentions from their input. User input consists of intent, noun phrases, and verb phrases. The NLP component uses rule-based semantic parsing \cite{hussain2019survey} to identify user expressions from the input. This component includes text input acquisition, text understanding, knowledge extraction, by using syntactic (syntax), and semantic analysis. The syntactic analysis includes defining the structure of the user expression based on grammatical analysis and generating labelled text as nouns, verbs, adjectives, etc. The semantic analysis rids the labelled text from structural ambiguity, lexical ambiguity, or both. We can compare the resultant ambiguity-free labelled text to predefined training phrases for all intents, to identify the required intent. Noun phrases can be divided into regular and proper nouns. Regular nouns can be matched to the predefined entities of the intent, while the proper nouns can be the names (identifier) of these entities. Verb phrases define the context and the current state of the intent flow. 
\subsubsection{DSL Model Construction} 
    This step uses the second layer discussed in Section \ref{IL}, to map the user input to the model concepts. The output is a DSL model (instance model) that represents user intention. The DSL follows a modular data structure similar to the JSON structure, which enables switching between the defined concepts. This makes it possible to query, update or even delete the defined model more flexibly and accessibly. Moreover, the constructed DSL enables the ability to trace the created instance model or generated code to the use case defined in the user conversation.

\subsubsection{Model Validation} The user intent captured by NLP is validated against a set of rules. These rules are based on the requirements of the abstract syntax of the meta-model. These form the criteria that user input must satisfy to guarantee that the desired output is complete, and there is no missing data. This is an important step since the user might not be a domain expert and does not have the required knowledge and skill to produce complete code.  

\subsection{Code Generation} The code of the smart contract is generated automatically from the constructed DSL model based on model-to-text transformation templates. A transformation template defines the rules to transform a DSL instance model into the targeted programming language that conforms to the correct syntax of the platform. 

\section{Implementation} \label{IMP}
Having illustrated our general approach to integrating chatbots with MDE frameworks, we turn next to the demonstration of an implementation of this approach using smart contract development as a case study. First, we discuss the meta-model used for modelling the smart contract. Then we present the chatbot implementation. Finally, the code generation for smart contracts is discussed.

\subsection{Smart Contract Case Study } \label{MR}

As a case study, we develop a chatbot to auto-generate smart contract codes based on model specifications. A smart contract is a computer code that is deployed in blockchain to enforce agreements when conditions are met \cite{buterin2014next,zou2019smart,zheng2020overview}. Nowadays, designing and developing smart contracts, in general, are non-trivial and require a high knowledge of the underlying blockchain platform architecture and consensus \cite{zou2019smart,zheng2020overview}. To tackle this issue, several research studies \cite{tran2018lorikeet,skotnica2020towards, sharifi2020symboleo, soavi2020contract} proposed the use of domain-specific model language to develop smart contracts. 

We have previously \cite{hamdaqa2020icontractml} proposed iContractML a  platform-independent reference model for smart contract design and development. Furthermore, we introduced iContractBot \cite{qasse2021icontractbot} a chatbot framework to generate smart contracts based on the reference model proposed in \cite{hamdaqa2020icontractml}. 
To show the feasibility of the proposed approach, we extend the iContractbot framework to include user input sensitization, model query and model validation. 

\subsection{Smart Contracts Meta-model}\label{subsec:dsl}
In this paper, we have adopted the meta-model of iContractML that we have proposed previously in  \cite{hamdaqa2020icontractml}. iContractML is a platform-independent reference model for smart contract design and development. The proposed smart contract reference model includes (SContract) a business logic that has a name and a target platform. The target platform implies the language in which the smart contract code will be generated. The framework presented here enables a user to generate code in three languages - Solidity, Hyperledger Composer and Microsoft Azure. 
The contract consists of three \emph{elements} which are asset, participant, and transaction.  Assets are tangible or intangible goods (e.g money, real estate, or vehicles) stored in the blockchain. Participants are contractual parties that have certain access rules (conditions) to execute transactions (actions) to change the state of the assets.
The elements in the smart contract are connected via \emph{relationships}. Relationships connect a transaction to either a participant or an asset. Thus, the framework allows the creation of various relationships in transactions. Parameters and relationships form the building blocks of the elements.
We have implemented the meta-model in \cite{hamdaqa2020icontractml} using text-based modelling in Xtext\cite{bettini2016implementing}. The reason for using textual modelling is that the user intents captured by the chatbot are in text form. 

\subsection{Intermediate Layer for Smart Contract}
After identifying the concepts of smart contract abstract syntax, we need to identify and generate the intermediate layer which will be used for mapping. To target a wider range of users, the framework must enable users to freely express their smart contract use cases with the terminology they know. For instance, depending on who the user is (e.g stakeholder, domain expert, or developer), the "participant" concept must map to a term they use (e.g. "contractual party", "role", or "state variable"). Also, different blockchain platforms use different terms to define the same smart contract concepts. For example, the contract method is referred to as a function in Ethereum and a transaction in Hyperledger Composer. This terminological mismatch might confuse developers and force them to learn multiple vocabularies for different blockchain platforms.
We generate a list of words that represent the synonyms of the concepts, their possible types, and actions that can perform on these concepts. Table \ref{tab:tab1} shows a sample of the generated synonyms for the smart contract meta-model concepts.
\begin{table}
\caption{Sample of the generated synonyms for the smart contract meta-model concepts}
\label{tab:tab1}
\centering
\begin{tabular}{|c|c|c|c|c|}
\hline
Concepts & Contract    & Participant & Asset          & Transaction \\ \hline
\multirow{3}{*}{\begin{tabular}[c]{@{}c@{}}NL \\ synonyms\end{tabular}} & Template & \begin{tabular}[c]{@{}c@{}}Contractual\\  Party\end{tabular} & Good & Method \\ \cline{2-5} 
         & Application & Role        & Struct         & Function    \\ \cline{2-5} 
         & Program     & Party       & State variable & Choice      \\ \hline
\end{tabular}%
\vspace{-5mm}
\end{table} 
\subsection{Chatbot Implementation}\label{subsec:chatbot}
\begin{figure} 
\resizebox{\linewidth}{!}{
\centerline{\includegraphics[scale = 0.4]{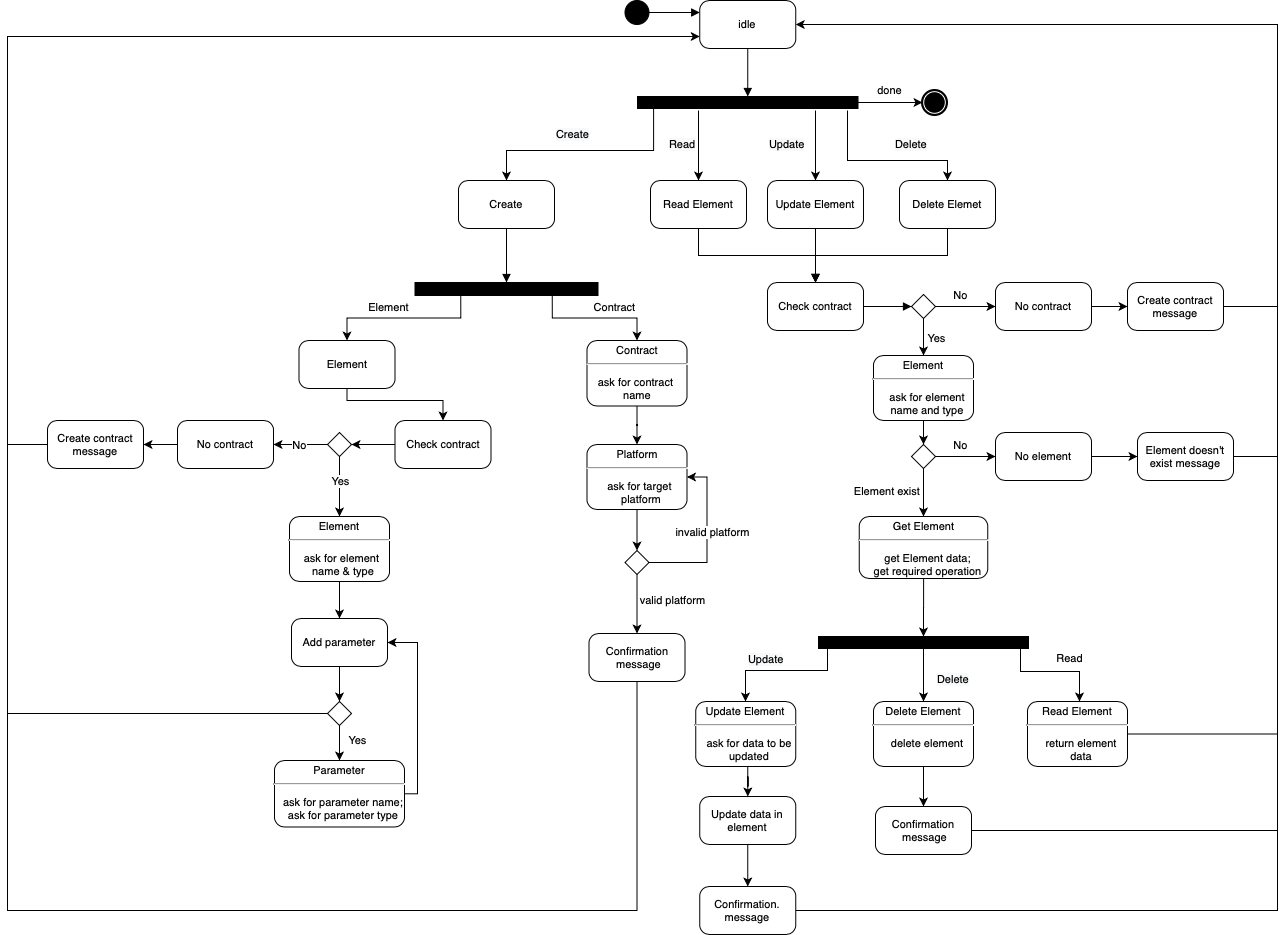}}}
\caption{Chatbot Intent Flow for Smart Contract Example}
\label{fig:INM}
\vspace{-5mm}
\end{figure}

In this paper, we adopted Xatkit \cite{daniel2020xatkit} as the bot framework to build the conversational bot, and to detect the user input. Xatkit \cite{daniel2020xatkit} is an open-source framework that helps to capture user intent and understand advanced natural language. The framework empowers building platform-independent chatbots \cite{daniel2020xatkit}. Xatkit is built on Java programming language, which makes it a perfect fit to integrate with modelling frameworks that are built on top of the Eclipse Modeling Framework. 
The framework was implemented to support the four operations (actions) and their possible synonyms: (i) Create; (ii) Edit; (iii) Delete, and (iv) Read. These four actions represent the basic model manipulation operations, and we have used them to identify the chatbot intent flow. The statechart in Figure \ref{fig:INM} shows the chatbot flow to create a smart contract use case, where an element can be an asset, participant or transaction. In all scenarios (model instance creation or query), a contract must first be defined, as illustrated in Figure \ref{fig:INM}.

\begin{figure}[t]
\centering
\includegraphics[scale =0.4]{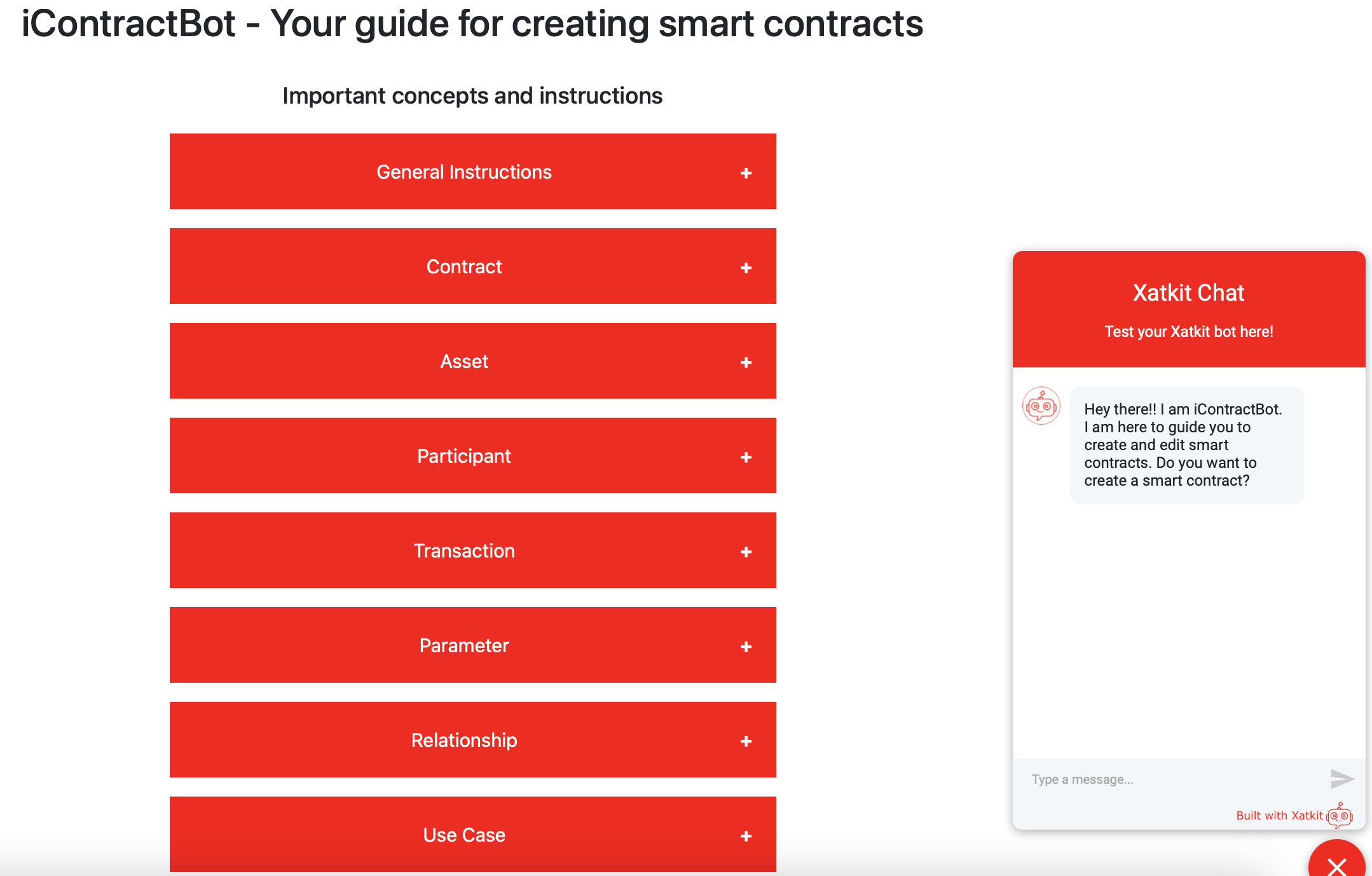}
\caption{An illustration of the chatbot}
\label{fig:botImp}
\vspace{-5mm}
\end{figure}

The chatbot performs various tasks according to the \emph{actions} specified by the user. This action definition improves chatbot performance greatly because there's a definite set of tasks that the chatbot must perform. This makes chatbot flow smoother, which in turn enhances entity detection. With the finite number of steps, a user finds it easier to direct the chatbot to perform various tasks. After being directed to perform a particular action, the chatbot asks the user for information needed for model completion. The chatbot tries to direct the user through a path that helps in the smooth creation of a complete smart contract model (the user does have to be fairly exhaustive in specifying what he/she intends to do). The Xatkit framework enables the deployment of the bot on multiple platforms including web pages, and social media platforms. In this paper, we have used React platform to deploy our chatbot on a web page. In the chatbot web page, we introduced the main components of the smart contract with examples to help any users to facilitate their interaction with the chatbot, as shown in Figure \ref{fig:botImp}. This is to ensure that people without any prior knowledge in blockchain or smart contracts have an idea of what exactly they are building. Moreover, within the chat, we introduced these components briefly to the user.
During the creation of the chatbot model, errors can emerge at three points: (i) while the chatbot switches between two elements; (ii) while the chatbot detects the actions and intents; and (iii) from the user's end while giving inputs. We address these failure points using several techniques, including input sanitization and input validation.
\subsubsection{DialogFlow for intent detection and entity extraction}
In the case of intent detection and entity extraction, Xatkit performs well up to a certain extent. Therefore, to enhance the performance of the chatbot, DialogFlow \cite{sabharwal2020introduction} has been integrated. By using DialogFlow, the chatbot is able to understand the numerous variants of the test sentences, which improves entity detection significantly. Dialogflow uses rule-based grammar matching \cite{sabharwal2020introduction} to detect user intent. It compares user input to the predefined training sentences for all intents, to identify the best matching intent.  Although Dialogflow improves the intent detection module, it can't detect phrases, where the user should provide meaningful sentences. For example, if the user wants to create a contract when the chatbot asks what they like to create, then the input should be \emph{"Create a contract"} or \emph{"I want to create a contract"}, not only \emph{"A contract"}. Thus, a user has to be elaborate while instructing the chatbot at each instance. If enough information is not provided, then the chatbot may generate incorrect codes (for instance if a user wants to create a contract and doesn't include the word create or any of its synonyms, the chatbot may end up considering an intent to edit an element) or the user will not be able to complete building the desired use case.

\subsubsection{Input Sanitization}
While editing, deleting, and reading elements, the user has to specify their names. If the user makes mistakes during this process, the chatbot won't be able to detect user intent correctly. As mentioned in Section \ref{AA}, we used Levenshtein Distance \cite{yujian2007normalized} to find the most similar string.  Dynamic programming is used to implement the Levenshtein Distance algorithm, which is much better optimized than a recursive approach. When a user asks to edit/delete/read an element, the chatbot obtains the edit distance between the entity extracted and all predefined entities. After finding the closest string, the chatbot modifies the input and confirms the modification with the user. This has also been for the datatype of the parameters. This ensures that the DSL generated is valid and the chatbot doesn't break during code generation (since we have predefined datatypes in the metamodel, a misspelt datatype would make the model invalid). This extra step makes the chat flow a little tedious but ensures accurate editing of the model.

\subsubsection{DSL Model Instance Construction}
A DSL has to be defined to interact as an intermediate layer between the user input and the smart contract code generation that will be used to generate the code. The DSL model instance consists of the user intent and the main concepts of the abstract syntax of the meta-model. The instance is generated by assigning the captured user intent throughout the conversation to the meta-model concepts in a structural way. From the DSL model instance, we generate the model instance of the meta-model. The structure of the DSL model instance uses a syntax similar to that of a JSON structure (it is not exactly the same as a JSON file). This makes it easy to switch between elements in the model, thus reducing the chance of errors made by the chatbot. For generating the DSL file from the chatbot, the chatbot stores the data in \emph{ArrayLists} of data structures, which were defined by us. Using such data structures, not only enhances the efficiency of the CRUD operations but also makes it easy for the chatbot to write to a DSL file. On the other hand, choosing a string data type within the chatbot as the structure for the DSL may result in a simpler process, however, it might also be error-prone. Furthermore, using the proposed structure will not affect the speed of the chatbot since the tasks being performed are computationally inexpensive. Hence, the usage of this data structure is justified. We have tried to design the DSL such that the components of the smart contracts are apparent to any user from the DSL. Any other syntax may also be used for the DSL if it provides huge benefits over the current structure of the DSL.
\subsubsection{Model Validation}
To ensure the validity of the model instance created by the user, we have implemented a set of validation rules in our framework. A sample of the validation rules are:
\begin{itemize}
    \item A target platform must be specified.
    \item A contract name must be specified.
    \item A user must specify the type of the asset.
    \item A unique identifier must be specified for an asset and participant.
    \item The user must specify a valid relationship that relates to an existing object (participant, asset, or transaction).
\end{itemize}
The model validation ensures that the generated DSL model instance is complete and accurate, which in results can be used to generate a correct complete code. 

\subsection{Code Generation}\label{subsec:codegen}
\begin{table*}[t]
  \centering
  \caption{Transformation Table Template}
  \begin{tabular}{ | m{2cm} | m{3.5cm}| m{3.5cm} | m{6cm} | }
    \hline
    Concept & Abstract Syntax & DSL Model Instance & Transformation Code Snippet (Ethereum) \\
    \hline
    Smart Contract / File &  \includegraphics[width=0.2\textwidth]{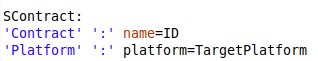} & \includegraphics[width=0.12\textwidth]{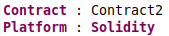} & \includegraphics[width=0.22\textwidth]{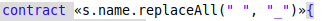} \\
    \hline
    Participant & \includegraphics[width=0.18\textwidth]{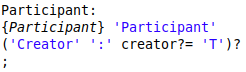} & \includegraphics[width=0.12\textwidth]{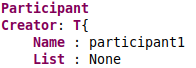} & \includegraphics[width=0.3\textwidth]{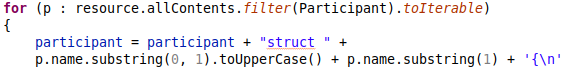} \\
    \hline
    Asset & \includegraphics[width=0.12\textwidth]{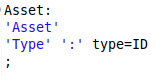} & \includegraphics[width=0.10\textwidth]{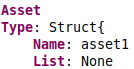} & \includegraphics[width=0.32\textwidth]{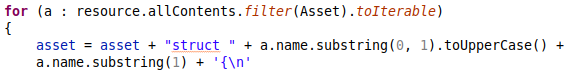} \\
    \hline
    Transaction & \includegraphics[width=0.2\textwidth]{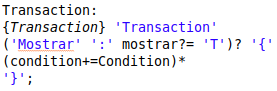} & 
    \includegraphics[width=0.13\textwidth]{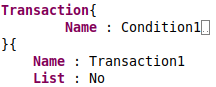} & \includegraphics[width=0.3\textwidth]{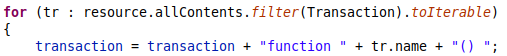} \\
    \hline
  \end{tabular}
  \label{tab:transformation-table}
  \vspace{-5mm}
\end{table*}
To generate the smart contract code from the DSL instance model generated by the chatbot,  we used Xtend\cite{bettini2016implementing} to write transformation templates for different blockchain platforms.  The transformation template takes as input the generated instance model and applies a set of transformation rules to generate a smart contract code for any of the three target platforms supported. The transformation templates used in this framework are based on the transformation templates of iContractML \cite{hamdaqa2020icontractml}.
To reduce the chances of errors, we made the design as modular as possible. This was done by creating a separate executable file for code generation which was run by the chatbot only at the end of the conversation. This separates the code generation part and DSL model construction part and ensures that any error in the model is rectified before it is fed to the code generation section.
Table \ref{tab:transformation-table} illustrates the transformation template code snippet for generating Ethereum code from the corresponding DSL instance model. 

All the source code for implementing the chatbot for smart contracts (including the transformation template, etc) are provided in the project repository \footnote{https://zenodo.org/record/5574618}.

\section{Evaluation} \label{EV}
In this section, we will evaluate the proposed chatbot framework for auto-generating smart contracts.  To evaluate the functionality of the chatbot and the correctness of the generated artifacts we have exploited three use cases in different fields.
Moreover, we have conducted a user experience survey to assess the usability of the chatbot. 

\subsection{Use Cases}

\begin{figure}
\centerline{\includegraphics[scale=0.52]{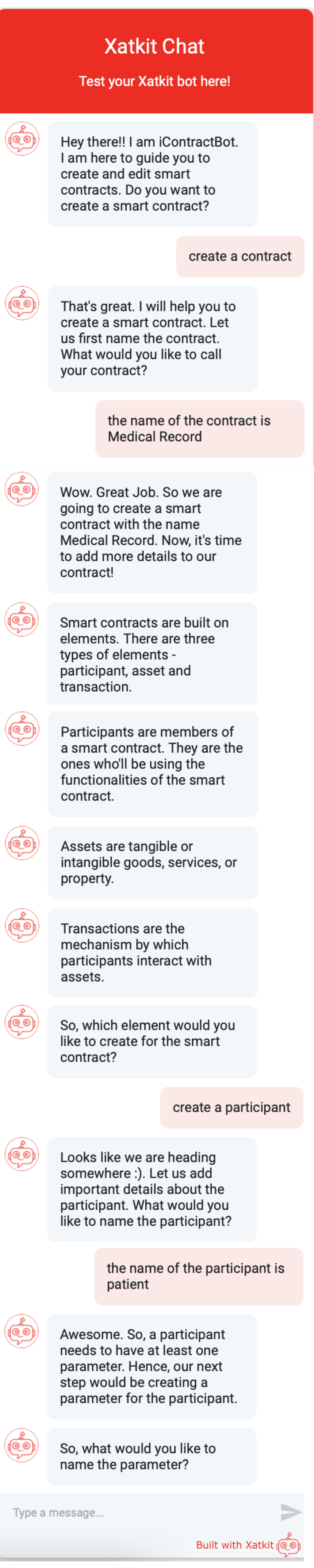}}
\caption{Medical Record Use Case Example}
\label{fig:CS}
\end{figure}

To evaluate the generated artifacts of the iContractBot, and the usage of the framework in different real-life fields, we have built models of three use cases from different applications including healthcare, education, and business.
Details of the use cases follow:
\begin{itemize}
\item Medical Records: a smart contract is used to store medical records for patients. The medical record consists of an id and the record owner. The patient can create the contract and also modify his/her details using the contact methods. 
    \item Digital Certificate: This smart contract verifies certificates obtained by students from educational institutions. It involves two participants, one asset, and two transactions. The asset is a "certificate" with three parameters: the hash, "verifiers", and "status" of the certificate. The two participants involved are the "issuer" and "verifier". The issuer issues the certificate with the Create Certificate transaction, while the verifier rejects or approves the issued certificate.
  \item Vehicle Auction: A smart contract is used to auction off vehicles. The vehicle is the key asset, with two parameters: "name" and "bid amount". There are two participants: "owner" and "bidder". The smart contract is created by the owner to auction off his/her vehicle. The bidder can place bids on the vehicles that they are interested in. The owner can use the "transfer" function to transfer ownership to the bidder.
\end{itemize}
The generated artifacts using the chatbot framework were deployed on their perspective blockchain platform, which did not result in any structural or syntax errors.
Figure \ref{fig:GC} illustrates the generated DSL model instance and the smart contract code for the medical record use case in Ethereum. A sample of the conversation to build the use case is shown in Figure \ref{fig:CS}
All the artifacts generated for the use cases are available in the project repository \footnote{https://zenodo.org/record/5574618}.

 \begin{figure}
\caption{Sample of the Generated Model Instance of DSL and Smart Contract Code in Ethereum}
\vspace{1mm}
\label{fig:GC}
\noindent\begin{minipage}{.48\textwidth}
\begin{lstlisting}[caption=Generated DSL Model Instance.]
Contract: MedicalRecord
Platform: Solidity
Participant
Creator: T {
Name: patient
List: Participant_List_None_0
Parameter {
Name: name
Type: String
 .... 
\end{lstlisting}
\end{minipage}\hfill
\begin{minipage}{.48\textwidth}
\begin{lstlisting}[caption=Smart Contract Code in Ethereum.]
pragma solidity >=0.4.22 <0.7.0;
contract MedicalRecord{
    constructor(){}
    struct Patient{
    bytes32 name;
    address patientAddress;} 
    struct Record{
    bytes32 id;
    bytes32 owner;} 
      ....}
\end{lstlisting}
\end{minipage}
\vspace{-5mm}
\end{figure}

\subsection{User Experience Survey}
\subsubsection{Design}
To evaluate the functionality and usability of the proposed chatbot, we conducted an online survey. We provided tutorials and video \footnote{https://youtu.be/JybbTgtat5M} to guide participants in using the chatbot. We also presented them with a basic overview of smart contracts and related concepts, accessible to non-technical users. We used the survey to answer three research questions:
\begin{itemize}
    \item RQ 1. What is the overall experience of the participants based on their background?
    \item RQ 2.  What are some current limitations and challenges with the chatbot?
    \item RQ 3. What are the possible improvements to improve the adaptability of the chatbot?
\end{itemize}

The survey questionnaire \footnote{https://forms.gle/2zHq6ktRoMVKCApK8} consists of four parts, which are:
\begin{itemize}
    \item Demographics: the questions in this part are related to the participant's background and experience.
    \item Pre-test: in this part, we have different questions based on the participant background. For developers, we aim to understand what tools that they normally use to develop their applications and how comfortable they are with these tools. For non-technical participants, we ask if they require any programming task in their work/daily life and how they accomplish this task. 
    \item Functionality and Usability: in this part, the questions are designed to answer the research questions. We have adopted The Unified Theory on Acceptance and Use of Technology (UTAUT), which was introduced by Venkatesh et al. \cite{venkatesh2003user}. From this model, we have considered six theories of technology adoption that fit with chatbots, which are performance expectancy, effort expectancy, attitude toward using technology, facilitating conditions, self-efficacy, and behavioral intention to use the framework. The list of the questions of this part is shown in Table \ref{tab:table3}.
    \item Post-test: In this part, we ask the participants about the overall experience and if they have any feedback regarding the chatbot. 
\end{itemize}
\begin{table}[]
\centering
\caption{List of questions included in the functionality and usability part}
\label{tab:table3}
\begin{tabular}{|c|c|c|}
\hline
Metric &
  ID &
  Question \\ \hline
\multirow{2}{*}{Functionality} &
  Q1 &
  \begin{tabular}[c]{@{}c@{}}Were you able to generate the smart \\ contract given in the test case?\end{tabular} \\ \cline{2-3} 
 &
  Q2 &
  \begin{tabular}[c]{@{}c@{}}How many times did the chatbot\\  respond in the wrong way?\end{tabular} \\ \hline
\begin{tabular}[c]{@{}c@{}}Performance\\ Expectancy\end{tabular} &
  Q3 &
  How useful do you think the chatbot is? \\ \hline
\multirow{2}{*}{\begin{tabular}[c]{@{}c@{}}Effort\\ Expectancy\end{tabular}} &
  Q4 &
  \begin{tabular}[c]{@{}c@{}}My interaction with the chatbot \\ is clear and understandable\end{tabular} \\ \cline{2-3} 
 &
  Q5 &
  \begin{tabular}[c]{@{}c@{}}What do you have to say about the amount \\ of information that had to be typed for generating the code?\end{tabular} \\ \hline
\multirow{2}{*}{\begin{tabular}[c]{@{}c@{}}Attitude toward \\ using technology\end{tabular}} &
  Q6 &
  \begin{tabular}[c]{@{}c@{}}How would you rate the chatbot in terms \\ of the interest instilled by the chatbot?\end{tabular} \\ \cline{2-3} 
 &
  Q7 &
  What were the reasons behind the chat being boring? \\ \hline
\multirow{4}{*}{\begin{tabular}[c]{@{}c@{}}Facilitating \\ Conditions\end{tabular}} &
  Q8 &
  I have the resources necessary to use the chatbot \\ \cline{2-3} 
 &
  Q9 &
  I found the video tutorial helpful \\ \cline{2-3} 
 &
  Q10 &
  \begin{tabular}[c]{@{}c@{}}How good was the chatbot in giving some \\ clarity about the working of smart contracts?\end{tabular} \\ \cline{2-3} 
 &
  Q11 &
  The chatbot is compatible with other platforms I use \\ \hline
Self-efficacy &
  Q12 &
  \begin{tabular}[c]{@{}c@{}}How many times did you refer to the \\ tutorial/ video in order to build the chatbot.\end{tabular} \\ \hline
\begin{tabular}[c]{@{}c@{}}Behavioral \\ intention to use \\ the system\end{tabular} &
  Q13 &
  I plan to use the chatbot in the future \\ \hline
\end{tabular}%
\vspace{-5mm}
\end{table}
\subsubsection{Survey Respondent Recruitment and Statistics}
In this paper, we aim to analyze the functionality and usability of the chatbot framework from different users' perspectives. We contacted potential participants in multiple ways. We broadcasted our survey in research survey forums and we sent emails to smart contract developers on GitHub. Moreover, we also asked our colleagues in the industry to help broadcast our survey to their friends and colleagues who may be interested to participate in our survey. From there, we got 46 responses. 
Out of the 46 respondents (20 females, 26 males), 7 hold a Master’s degree, and 27 have a Bachelor’s degree, where 12 of them are pursuing their Master’s degree. 10 of the participants are still pursuing their Bachelor's degree, while 2 of the participants are doctoral students. Only 19.6\% of the participants are familiar with MDE.

To better understand the participants' experience we divided the survey respondents into different demographic groups: developers with blockchain background, general developers, and non-programmers.
Among these 46 participants, 34.78\% are developers with blockchain background, 36.96\% are general developers only and 28.26\% non-programmers. 
69.7\% of the developers are familiar with  1 to 5 programming languages. On the other hand 27.3\%  of the developers know between 5 to 10 programming languages,  while 3\% know only one programming language. The programming language that most of the developers are familiar with are python, java, Matlab, Go language, etc.  
48.5\% of the developers feel comfortable with a new programming language within 2 to 4 weeks, while 42.4\% of them get familiar with a new programming language within less than 2 weeks.

62.5\% of the developers with blockchain background possess in-depth knowledge about smart contracts and cryptocurrency. These developers faced many challenges with current smart contract languages, namely a lack of documentation and support (leading to errors in their code), and the completely new syntax that differs based on the underlying blockchain platform. 
\subsection{Findings}
In this section, we discuss the result of evaluating the functionality and usability of the chatbot. Table \ref{tab:S-table} summarizes the results of the survey in the different metrics.
\subsubsection{Functionality}
In terms of functionality, we asked the participants to follow a predefined case in the tutorial. The main functionality of the chatbot is to generate smart contract code from the specification of the user. 78.3\% were able to generate the final code of the smart contract. 21.7\% were not able to generate the contract code but generated the DSL. 
We also evaluated the number of times that the chatbot responded incorrectly or was not able to detect the user's intention through the conversation. The chatbot responded incorrectly more than once but less than five times for 45.7\% of the participants. 28.3\% of the respondents countered chatbot errors more than five times but less than ten times. Only a few users (10.9\%) faced more than ten chatbot errors, while the chatbot responded incorrectly only once for 15.2\% participants through the conversation.

\subsubsection{Performance Expectancy}
This refers to the extent to which the participants believe the system is useful. We asked the participants how useful they think the chatbot was before using it (as an idea) and afterwards. The results show that 97.8\% of the participants thought that the idea of generating codes/smart contracts from the conversation is useful and helpful (evaluated 5 and above, where 7 is very helpful/useful). Hence, the idea is appealing to a large percentage of users. After using the chatbot, we have asked again the participants to evaluate the usefulness of the chatbot.  93.5\% of the participants still evaluated the chatbot as useful (evaluated 5 and above), where only 4.3\% of the participants rated the chatbot as not useful. 

\subsubsection{Effort Expectancy} This refers to the effort required from the users to use the system, in our case the chatbot. To assess this we evaluated how clear and understandable the interaction with the chatbot was. The results show that 82.6\% of users scored it from 5 to 7, where 7 indicates that the interaction was very clear and understandable. 10.9\% of the participants gave it a neutral score of 4. Only three participants (6.5\%) indicated that the interaction with the chatbot was not clear. 
We also asked users about their impression of the amount of data they were required to type (i.e. conversation length) to generate the code. 47.8\% of the users indicated that the amount of typing was acceptable (evaluated 1 or 2, where one is acceptable), while on the other hand, 37\% thought there was too much typing (evaluated at 4 or 5, where 5 is too much). 15.2\% were neutral and indicated that the amount was not too much/just right.

\subsubsection{Attitude toward using technology} This is a theoretical measure of participant attitudes using a tool/system. We asked participants to rate the chatbot in terms of how interesting they thought it was to interact with the chatbot. 80.4\% rated the chatbot between the scores 1 to 3 (where 1 is very interesting), while 15.2\% rated the conversation with the chatbot as boring (the scores are 5 or 6). Issues that affected the participants' attitudes included typing a lot, repetitive messages from the chatbot, etc.

\subsubsection{Facilitating Condition} This refers to the extent that the participant can acquire the required skills and knowledge from the technical infrastructure to use the tool. In this paper, we measured if we provided users with enough resources and tutorials to help them use the chatbot correctly. 97.8\% indicated that the platform provided the required resources to use the chatbot. All the participants indicated that they found the video tutorial very helpful. However, 
69.6\% evaluated that the chatbot was compatible with other platforms they use. This is due to that some participants faced an issue with installing the chatbot in their machines as the Xatkit framework doesn't support windows operating systems.  
We also evaluated how much these resources contributed to participants' knowledge. The results showed that 67.4\% of the participants did not know about smart contracts before using the chatbot, where 32.6\% of them gained a lot of knowledge of this concept, while 28.2\% got some clarity regarding the smart contract concept after using the chatbot. 

32.6\% already had previous knowledge of smart contracts. Therefore, a tool, that can generate codes from natural language, can be a medium of learning and thus can bridge the gaps between various fields.

\subsubsection{Self-efficacy} This measures the user's ability to use the chatbot without needing support from a technical person or tutorial. 13\% of the participants referred to the tutorial only once while using the chatbot. 34.8\% required tutorial help between one to five times during the conversation, while 41.3\% referred to a chatbot more than 5 times but less than 10 times. Some of the participants (10.9\%) were not able to interact with the chatbot easily and referred to the provided tutorials more than ten times. One possible reason for the less usage of the tutorial is the video tutorial. This shows that a basic interactive introduction(like video) to technology can greatly reduce the requirement of technical support.

\subsubsection{Behavioral Intention to use the system} We asked the participants if, based on their experience with the system, they might use the chatbot in the future. The question was presented on a scale from 1 to 7, where 1 represents very unlikely and 7 represents very likely. We got a positive response from 56.52\% that they would likely use the chatbot in the future, while 30.44\% were not likely to use the chatbot again and the rest 13.04\% weren't sure about using it again. After analyzing the user responses, most of the participants who said they were unlikely to use the system again were non-programmers who never programmed and didn't or rarely had the intention to program. 
\begin{figure} [t]
\centering
\includegraphics[scale = 0.5]{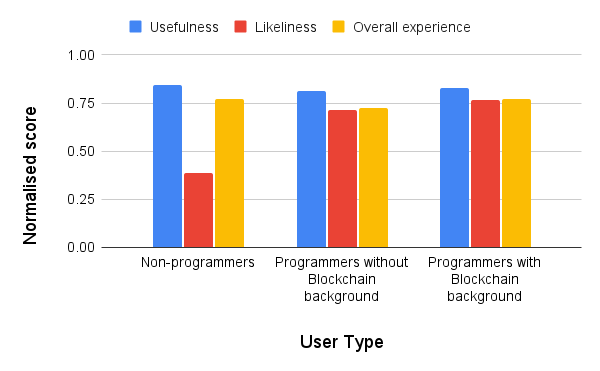}
\caption{Metrics to understand expectations and overall experience of the users}
\label{fig:overallResults}
\vspace{-5mm}
\end{figure}
\subsection{Discussion}
In this section, we will discuss the overall user experience with the chatbot and the factors that affected or improved it. These factors were extracted from participants' comments and suggestions in the survey. 
\subsubsection{What is the overall experience of the participants based on their background?}
Figure \ref{fig:overallResults} depicts the overall experience of the participants with the chatbot based on their background. We have presented three metrics - usefulness, likeliness (of using the bot again) and overall experience. 
For measuring usefulness, we studied the performance expectancy scores and for the likeliness, we considered the behavioral intention. In the case of the overall experience, we asked the users to rate the chatbot on a scale of 1-10. 
The values depicted by the bar graphs is the normalised average score of that metric in a particular category. The normalised score ($NS_{ij}$) is given by:

\begin{align*}
    NS_{ij} = \frac{AVG(Scores_i)}{Max_j}
\end{align*}
where,
\begin{align*}
    AVG(Scores_i) = \frac{(\text{\textit{Sum of scores of users in category i}})}{N_i}
\end{align*}
, $i\in [\text{\textit{Non-programmers, Programmers without blockchain}}$ \\ $\text{\textit{background, Programmers with blockchain \newline background}}]$, $j\in [\text{\textit{Usefulness, Likeliness, Overall experience}}]$, $Max_j$ is the maximum possible score that can be provided by a user in the metric $j$ and $N_i$ is the number of participants in the category $i$.
The results show that the normalised score of the overall experience is similar for the three different user types. However, for the likeliness metric, it is lower for non-programmers compared to the other two participants types. This is expected as most of the non-programmers participants didn’t require programming in their work or daily life. It should be noted that the people of all the three categories had similar overall experiences and find the chatbot almost equally useful. Moreover, in general, the usefulness score is more than the likeliness and the overall experience metrics. Both of the above-mentioned derivations from the figure indicate that such a tool would prove to be helpful for people of all backgrounds.

Furthermore, we asked the participants to identify the challenges faced that affected their experience. Moreover, we have also obtained the possible improvements from them for future enhancement of the chatbot.

\begin{table*}[t]
\centering
\scriptsize
\caption{Summary of the Survey Results}
\label{tab:S-table}
\begin{tabular}{|c|c|c|}
\hline
Metric &
  Question &
  Results \\ \hline
\multirow{2}{*}{Functionality} &
  Q1 &
  \begin{tabular}[c]{@{}c@{}}78.3\% generated the code\\ 21.7\% couldn't generate the code\end{tabular} \\ \cline{2-3} 
 &
  Q2 &
  \begin{tabular}[c]{@{}c@{}}The chatbot responded incorrectly:\\ - once for 15.2\% \\ - between one and five times for 45.7\% \\ - between five and ten times for 28.3\%  - more than ten times for 10.9\%\end{tabular} \\ \hline
\begin{tabular}[c]{@{}c@{}}Performance\\ Expectancy\end{tabular} &
  Q3 &
  93.5\% thinks chatbot is useful \\ \hline
\multirow{2}{*}{\begin{tabular}[c]{@{}c@{}}Effort\\ Expectancy\end{tabular}} &
  Q4 &
  \begin{tabular}[c]{@{}c@{}}The interaction with the chatbot:\\ - was clear for 82.6\%-was not clear for 6.5\%\end{tabular} \\ \cline{2-3} 
 &
  Q5 &
  \begin{tabular}[c]{@{}c@{}}The typing amount:\\ -  is acceptable for 47.8\%\\ - is too much for 37\% \end{tabular} \\ \hline
\multirow{2}{*}{\begin{tabular}[c]{@{}c@{}}Attitude toward\\ using technology\end{tabular}} &
  Q6 &
  \begin{tabular}[c]{@{}c@{}}80.4\% rated the chatbot as interesting\\ 15.2\% rated conversation is boring\end{tabular} \\ \cline{2-3} 
 &
  Q7 &
  \begin{tabular}[c]{@{}c@{}}The chatbot is boring due to:\\ - There is a lot of typing\\ - Repetitive messages\end{tabular} \\ \hline
\multirow{4}{*}{\begin{tabular}[c]{@{}c@{}}Facilitating\\ Conditions\end{tabular}} &
  Q8 &
  \begin{tabular}[c]{@{}c@{}}97.8\% indicated that the platform provides \\ the required resources to use the chatbot.\end{tabular} \\ \cline{2-3} 
 &
  Q9 &
  All the participants indicated that the video tutorial was helpful \\ \cline{2-3} 
 &
  Q10 &
  69.6\% indicated that the chatbot is compatible with other platforms they use \\ \cline{2-3} 
 &
  Q11 &
  \begin{tabular}[c]{@{}c@{}}32.6\% already knew about smart contracts\\ 32.6\% gained a lot of knowledge in smart contract\\ 28.2\% got some clarity regarding smart contracts\end{tabular} \\ \hline
Self-efficacy &
  Q12 &
  \begin{tabular}[c]{@{}c@{}}The participant referred to the tutorial:\\ - once for 13\% \\ - between one and five times for 34.8\% \\ - between five and ten times for  41.3\% \\ - more than ten times for 10.9\%\end{tabular} \\ \hline
\begin{tabular}[c]{@{}c@{}}Behavioral\\ intention to use the system\end{tabular} &
  Q13 &
  \begin{tabular}[c]{@{}c@{}}56.52\% will likely use the chatbot in the future\\ 30.44\% will unlikely use the chatbot\end{tabular} \\ \hline
\end{tabular}%
\vspace{-5mm}
\end{table*}

\subsubsection{What are the current limitations and challenges with the chatbot?} \label{lim} 
    
Most of the chatbot's errors resulted from user input errors, limitations in detecting the user intention, or technical questions\textbackslash responses that may have frustrated the user. Some user input errors resulted in failure to generate the smart contract codes in some cases, for example referring to non-existent variables such as participant, asset, or transaction. The chatbot didn't notify the users of these errors, which added some confusion. 
Limitations in input detection resulted from the sensitive structure of sentences that the NLP engine expected. Moreover, the NLP engine sometimes requires a full sentence, which increases the amount of user effort required to define a use case. This has caused some frustration and negatively affected the user experience. 

One of the main limitations that many participants pointed out was the amount of typing required to create the smart contract. This increased the likelihood of error and made the chatbot more complicated. Furthermore, some minor issues such as the level of detail of the chatbot questions/responses, or the long messages from the chatbot decreased the clarity and understandability of the interaction with the chatbot. 

\subsubsection{What are the possible improvements to improve the adaptability of the chatbot?} \label{improv} 

One important way to improve user experience and decrease input error would be to reduce the amount of typing the system requires. This may include providing template contracts, adding an auto-completion mechanism, and implementing drop-down lists for choosing from a set of constructed sentences. 
It would also be helpful to provide an intuitive way to track progress as the contract is written, so the user remains oriented within the larger process. A side-by-side GUI visualization might make that progress clear. Moreover, to avoid user input errors, an advanced input sanitization algorithm must be used to detect possible spelling mistakes or detect when the user modifies or uses non-existing variables. Furthermore, improving the NLP algorithm used and increasing the corpus of training sentences will help and boost the usability of the chatbot. 
To make the chatbot more accessible to non-technical users, the level of questions/responses of the chatbot should be simpler, and if that is not possible, it should be defined in simple terms as a note.

\section{Limitations of the Research} \label{LIMR}
The main goal of this paper is to study and investigate the use of chatbots in automatic code generation. However, we are still in the initial stages to achieve this goal, as the current approach and implemented chatbot have some limitations. 

In this paper, we have used smart contracts as a case study to show the feasibility of the approach. There is a need to investigate if the current approach can be extended in other software development domains such as mobile applications, web development, etc. Moreover, the current approach is not fully automated, especially the chatbot development from an existing conceptual meta-model. This includes automatically generating the intermediate layer as well as the chatbot intent flow from a meta-model. In the smart contract case study, we have developed the iContractbot manually from the reference model of iContractML. This will facilitate the process of adopting this approach in many software development fields. 

There are some limitations and challenges in the current implementation of the chatbot, which is discussed in Section \ref{lim}. These challenges include limitations in input detection, and identifying user errors. The chatbot requires some modifications which are discussed in Section \ref{improv} to enhance the usability of the chatbot. Nevertheless, there is a need for further user experience study with larger samples in different fields to identify the limitations of the proposed approach as well as the implemented chatbot. This may include comparing between different interfaces (chatbot, graphical or mix approach) and identify the best approach to target a wider range of users.

\section{Threat to Validity} \label{TH}
In this paper, we used smart contract development as a running example to automatically generate code from conversational syntax. A threat to the external validity of our research is whether or not our selected running example is general enough for our approach to be extended to other traditional domains of software development such as web development, object-oriented programming languages, Internet of things (IoT), etc. We maintain that smart contract development is complex and there are not many experienced developers in this field \cite{zou2019smart}, which serves to demonstrate the motivation of this work. However, there is still a need to investigate the other software development domains to assess the generalizability of the proposed approach.  

Another threat to validity is in the evaluation using case studies.  The purpose of the use cases is to demonstrate that the generated artifacts do not have any syntax errors.  However, this might be prone to researcher bias as we might select an application area that serves the iContractBot framework. To counter this we have chosen three use cases from different fields (e.g healthcare, education, business,etc.) that are popular use cases in the blockchain domain \cite{swan2015blockchain}. This way we mitigate bias toward certain fields or applications.

We conducted a survey with 46 participants to evaluate the implemented chatbot framework. A threat to validity might have been that we missed potential users who might have assessed our solution differently from those in our study. To reduce this threat, we targeted participants from different backgrounds with different levels of experience (smart contract developers, general developers, non-programmers). This diversity of backgrounds helped us reflect on real-world conditions of smart contract development. Nevertheless, the number of participants was small, and we need to conduct the survey on a larger scale for better feedback, which is the next stage of our project.  It is also possible that survey respondents may have provided biased answers, based on what we want to hear due to several reasons. To help in obtaining unbiased answers, we allowed the respondents to be anonymous, where they can share if they want their email address.

\section{Related Work} \label{RW}
This section presents the related work of using chatbot with MDE whether it was to generate conceptual models, query models, or developing chatbots. Furthermore, we present the related work of auto-generating smart contract artifacts from chatbots or NLP. 

Pérez-Soler et al. \cite{perez2019flexible} investigated the usage of NL with DSL and introduced an approach to automate the generation of chatbot interface from models in \cite{perez2019towards}. This work was evaluated using a real case study of the existing modelling framework. 

In another work, Pérez-Soler et al. \cite{perez2020towards} proposed the use of chatbots (conversational syntax) to query existing conceptual models. Pérez-Soler et al. \cite{perez2020towards} used the Xatkit framework to integrate chatbots with EMF-based meta models. 

Conga  \cite{perez2020model}, an MDE approach, was proposed to automatically develop chatbots and generate their codes in two platforms Dialogflow and Rasa. Conga also provides a recommendation tool to recommend the best chatbot platform to use based on the system specification. 

Ed-douibi et al. \cite{ed2020model} proposed an MDE approach to generate chatbots for Open Data web APIs queries. In this work, the Xatkit framework was used to generate chatbots that support both direct and guided queries.

Table \ref{tab:tab2} summarizes the related work and compares it with our approach. In general, there are few works done in the field of chatbot and MDE. Most of the discussed related work focuses on generating, querying models, or developing chatbots, while in our approach, we focused on querying and creating instances from models. Furthermore, we applied our approach to auto-generating codes from chatbots, in particular smart contract artifacts. There is also related research work in the field of using NLP to domain modeling such as in \cite{saini2020domobot, ibrahim2010class, robeer2016automated}. However, these works focus on analysing text (description) instead of interactive conversation as in our case.

\begin{table}[]
\caption{Related Works Comparison}
\label{tab:tab2}
\resizebox{\columnwidth}{!}{%
\begin{tabular}{|c|c|c|c|c|c|}
\hline
\begin{tabular}[c]{@{}c@{}}Related \\ Work\end{tabular} &
  \begin{tabular}[c]{@{}c@{}}Generating \\ Conceptual Model\end{tabular} &
  \begin{tabular}[c]{@{}c@{}}Query\\  Model\end{tabular} &
  \begin{tabular}[c]{@{}c@{}}Create \\ Model Instances\end{tabular} &
  \begin{tabular}[c]{@{}c@{}}Chatbot \\ Development\end{tabular} &
  \begin{tabular}[c]{@{}c@{}}Evaluation\\  Method\end{tabular} \\ \hline
\cite{perez2019towards}  & \checkmark &   &   & \checkmark & Case Studies                                                                                  \\ \hline
\cite{perez2019flexible} & \checkmark &   &   &   & Case Studies                                                                                  \\ \hline
\cite{perez2020model}    &   &   &   & \checkmark & Case Studies                                                                                  \\ \hline
\cite{perez2020towards}  &   & \checkmark &   &   & Proof of concept                                                                              \\ \hline
\cite{ed2020model}       &   &   &   & \checkmark & \_                                                                                             \\ \hline
\begin{tabular}[c]{@{}c@{}}Our \\ approach\end{tabular}             &   & \checkmark & \checkmark &   & \begin{tabular}[c]{@{}c@{}}Case Studies\\ Functionality and \\ Usuability Survey\end{tabular} \\ \hline
\end{tabular}
} \vspace{-5mm}
\end{table}

\section{Conclusion} \label{CON}
In this paper, we investigated the use of chatbots as the concrete syntax for modeling frameworks, Illustrate requirements, specify design, and auto-generate codes. We proposed an approach for auto-generating platform-independent codes, in particular smart contract codes, from conversational syntax. Moreover, we showcased our methodology with smart contract development and implemented a chatbot framework, iContractBot, for modeling and developing smart contracts. Furthermore, we evaluated the chatbot framework in terms of its usability, and functionality based on the user experience study. The results showed that the overall user satisfaction depends on the participant's background. However, 79\% of the participants in all groups had an above-average overall experience in using the framework.  

This work is considered as the initial step to bridge the gap between users and software development. The current approach and implementation require further improvements to achieve the optimal goal. In future work, we plan to extend the approach to other case studies in software development such as mobile development, IoT, etc. We will also conduct an in-depth empirical study, evaluating and comparing graphical and conversational interfaces as the concrete syntax for modelling frameworks. Moreover, we will examine the various improvements suggested by the participants in our current survey, which include creating hybrid conversational-graphical interfaces, providing template contracts, adding auto-completion mechanisms, etc. 

\bibliographystyle{unsrtnat}
\bibliography{bibliography.bib}

\end{document}